\documentstyle[12pt,epsfig,psfrag,wrapfig]{article}
\newcommand{\be}{\begin{equation}}
\newcommand{\ee}{\end{equation}}
\newcommand{\ba}{\begin{eqnarray}}
\newcommand{\ea}{\end{eqnarray}}
\newcommand{\Pnperp}{{\bf P}_{\!\!\!\perp\Nu}}
\newcommand{\Phperp}{{\bf P}_{\!\!\!\perp h}}
\newcommand{\Mn}{M_{\mbox{\tiny N}}}
\newcommand{\Nu}{  {\mbox{\tiny N}}}
\newcommand{\di}{ {\rm d} }
\newcommand{\eqdef}{\:\stackrel{\rm def.}{=}\:}
\newcommand{\la}{\langle}
\newcommand{\ra}{\rangle}
\newcommand{\athree}[3]{$\renewcommand{\arraystretch}{0.9}
            \begin{array}{l} #1 \\  #2 \\ #3 \end{array}$}

\begin{document}
\title{Azimuthal asymmetry in electro-production of neutral pions in
semi-inclusive DIS}
\author{A.~V.~Efremov$^a$\thanks{Partially supported by RFBR grant
  00-02-16696 and INTAS grant 1A-587.}, K.~Goeke$^b$, P.~Schweitzer$^{b}$ \\
  \footnotesize\it
  $^a$ Joint Institute for Nuclear Research, Dubna, 141980 Russia\\
  \footnotesize\it $^b$ Institute for Theoretical Physics II, Ruhr University
  Bochum, Germany}
\date{}
\maketitle
\vspace{-9cm}\begin{flushright} RUB/TP2-08/01\end{flushright}\vspace{7cm}

\begin{abstract}
\noindent
Recently HERMES has observed an azimuthal asymmetry $A_{UL}$ in
electro-production of neutral pions in semi-inclusive deep-inelastic
scattering of unpolarized positrons off longitudinally polarized protons.
This asymmetry (like those observed in the production of charged pions)
is well reproduced theoretically by using
the non-perturbative calculation of the proton transversity distribution
$h_1^a$ in the effective chiral quark-soliton model combined with
experimental DELPHI-data on the new T-odd Collins
fragmentation function $H_1^\perp$.
There are no free, adjustable parameters in the analysis. Using
the $z_h$-dependence of the HERMES azimuthal asymmetry and the calculated
transversity distributions the $z_h$-dependence of the Collins fragmentation
function is obtained.
The value obtained from HERMES data is consistent with the DELPHI result,
even though these results refer to different scales.
\end{abstract}

\section{Introduction}

Recently a large azimuthal asymmetry has been observed by HERMES in the
electro-production of neutral pions in semi inclusive deep-inelastic
scattering (SIDIS) of unpolarized positrons off longitudinally polarized
protons \cite{hermes-pi0}.
A similarly large azimuthal asymmetry in the production of $\pi^+$ has been
observed before, while no such azimuthal asymmetry was found in the
production of $\pi^-$ \cite{hermes}.
Azimuthal asymmetries were also observed in SIDIS off transversely
polarized protons at SMC \cite{bravardis99}.
These asymmetries contain information on
the proton transversity distributions
$h_1^a(x)$ and on the Collins fragmentation function
$H_1^{\perp a}(z_h)$ \footnote{
    We use the notation of the
    Ref. \cite{muldt,muldz,Mulders:1996dh}.}.
The transversity distribution function $h_1^a(x)$ describes the
distribution of transversely polarized quarks of flavour $a$
in the nucleon \cite{transversity}.
The T-odd fragmentation function $H_1^{\perp a}(z_h)$ describes the
left-right asymmetry in
fragmentation of transversely polarized quarks of flavour $a$
into a hadron \cite{muldt,muldz,Mulders:1996dh,collins,hand}
(the so-called "Collins asymmetry").
Both $H_1^{\perp a}(z_h)$ and $h_1^a(x)$ are  twist-2, chirally odd,
and not known experimentally.
Only in the last years experimental indications to the T-odd
fragmentation function $H_1^{\perp a}(z_h)$
in $e^+e^-$-annihilation
have appeared \cite{todd,czjp99},
while the HERMES and SMC experiments \cite{hermes-pi0,hermes,bravardis99}
can be viewed as the very first experimental indications to $h_1^a(x)$.

Here we will explain the azimuthal asymmetry in $\pi^0$ production
\cite{hermes-pi0} by using information on $H_1^{\perp}$ from DELPHI
\cite{todd,czjp99} and the predictions from the chiral quark-soliton model
($\chi$QSM) for the transversity distribution $h_1^a(x)$ \cite{h1-model}.
Our analysis is free of any adjustable parameters.
In this way the azimuthal asymmetries for $\pi^\pm$
\cite{hermes,bravardis99} have been explained in Ref. \cite{Efremov:2000za}.
We recalculate them using a bit more exact experimental cuts.
In Ref. \cite{hermes-pi0} the data is compared to results of a similar
analysis, which is based on the approach of Ref. \cite{Oganessyan} and
which, however, makes use of adjustable parameters and certain assumptions
about $h_1^a(x)$.

In order to use information from DELPHI on $H_1^\perp$, we have to assume
that $\la H_1^\perp\ra/\la D_1\ra$, the ratio of the T-odd to the usual
fragmentation function (averaged over $z_h$ and over flavours),
varies little with scale.
We will investigate whether this assumption is justified.
For that we will use the prediction of $h_1^a(x)$ from $\chi$QSM to extract
$H_1^\perp(z_h)$ from $z_h$-dependence of HERMES data.
We will show that the results for $\la H_1^\perp\ra/\la D_1\ra$ from
HERMES \cite{hermes-pi0,hermes}, SMC \cite{bravardis99} and
DELPHI \cite{todd,czjp99} are consistent with each other.

\section{Ingredients for analysis:
         \boldmath{$h_1$} and \boldmath{$H_1^\perp$}}

	\begin{wrapfigure}{HR}{5.5cm}
   	\psfrag{u}{\boldmath\footnotesize ${\rm u}$}
   	\psfrag{d}{\boldmath\footnotesize ${\rm d}$}
   	\psfrag{ub}{\boldmath\footnotesize $\bar{\rm u}$}
       	\psfrag{db}{\boldmath\footnotesize $\bar{\rm d}$}
   	\mbox{\epsfig{figure=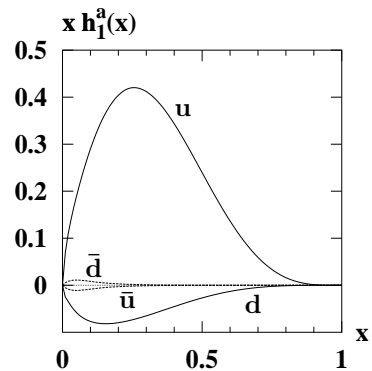,width=5.5cm,height=5.5cm}}
   	\caption{\footnotesize\sl
     	  The chiral quark-soliton model prediction for the proton
          $x h_1^a(x)$ vs. $x$ at the scale $Q^2=4\,{\rm GeV}^2$.
          The $u$-quark dominates the proton transversity distribution.}
	\end{wrapfigure}
\paragraph{Transversity distribution function \boldmath{$h_1$}.}
We will take the predictions of the chiral quark-soliton model
($\chi$QSM) as input for $h^a_1(x)$ \cite{h1-model}.

The $\chi$QSM is a quantum field-theoretical relativistic model
with explicit quark and antiquark degrees of freedom. This allows an
unambiguous identification of quark as well as antiquark distributions
in the nucleon.
Due to its field-theoretical nature the quark and antiquark distribution
functions obtained in this model satisfy all general QCD requirements:
positivity, sum rules, inequalities, etc \cite{DPPPW96}.
The model results for the unpolarized quark and antiquark distribution
function $f_1^a(x)$ and for the helicity quark distribution function
$g_1^a(x)$ agree within (10 - 20)\%
with phenomenological parametrizations.
This encourages confidence in the model predictions for $h_1^a(x)$.
In Fig. 1 the results of the model are shown at the average
$Q^2=4\,{\rm GeV}^2$ close to the HERMES experiment.

The application of the model results has yet another advantage.
When using the model results for twist-2 parton distributions it is
consequent to neglect systematically twist-3 distributions for the
following reason.
The $\chi$QSM has been derived from the instanton model of the QCD vacuum,
and in the latter nucleon matrix elements of twist-3 operators are
suppressed with respect to the leading twist-2 \cite{Diakonov:1996qy}.
In the case of the twist-3 distribution $\widetilde{h}_L^a(x)$ this
has been shown explicitly in Ref. \cite{Dressler:2000hc}.

\paragraph{The T-odd fragmentation function \boldmath{$H_1^\perp$}.}
The Collins fragmentation function $H_1^\perp(z_h,{\bf k}^2_\perp)$
describes a left--right asymmetry in the fragmentation of a transversely
polarized quark with spin {\boldmath$\sigma$} and momentum
${\bf k}= ({\bf k}_L,\,{\bf k}_\perp)$ into a hadron with momentum
${\bf P}_{\!h}=-z_h{\bf k}$: the relevant structure is
$H_1^\perp(z_h,{\bf k}^2_\perp)\;
 \mbox{\boldmath$\sigma$}({\mathbf k}\times
 {\mathbf P}_{\!\!\!\perp h})/(|{\bf k}|\la P_{\!\!\perp h}\ra)$.
Here $\la P_{\!\!\perp h}\ra$ is the average transverse momentum of the
final hadron\footnote{
    Notice the different normalization factor compared to
    \cite{muldt,muldz,Mulders:1996dh}, $\la P_{h\perp}\ra$
    instead of $M_h$.}.

This fragmentation function is responsible for a specific
azimuthal asymmetry of a hadron in a jet around the axis
in direction of the second hadron in the opposite jet.
This asymmetry was measured using the DELPHI data collection \cite{todd}.
For the leading particles in each jet of two-jet events,
averaged over quark flavours (assuming
$H_1^{\perp}=\sum_h H_1^{\perp\, q/h}$ is flavour independent), the
most reliable value of the analyzing power is given by
\be\label{apower}
   	\left|{\la H_1^{\perp}\ra\over\la D_1\ra}\right| =(6.3\pm 2.0)\% \ee
with presumably large systematic errors\footnote{
    A similar value was also obtained from the pion asymmetry
    in inclusive $pp$-scattering \cite{bogl99}.}.
The result Eq.(\ref{apower}) refers to the scale $M_Z^2$
and to an average over ${\bf k}_\perp$ and over
$z_h$ with $\la z_h\ra\simeq 0.4$ \cite{todd}.

\section{The HERMES experiment for {\boldmath$A_{UL}$}}

	\begin{wrapfigure}{HR}{6.5cm}
	\mbox{\epsfig{figure=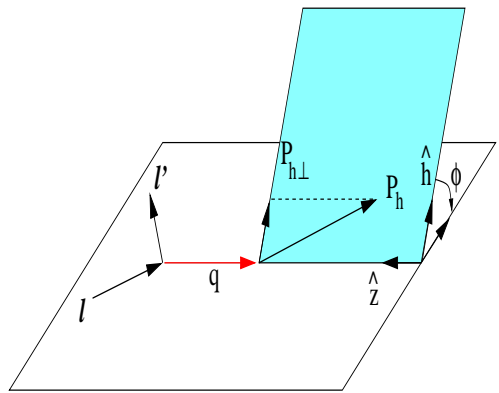,width=6.5cm,height=5.5cm}}
    	\caption{\footnotesize\sl
    	  Kinematics of the process $lp\rightarrow l'\pi X$
    	  in the lab frame. In the HERMES experiment
    	  the lepton $l$ is a positron.}
	\end{wrapfigure}
In the HERMES experiment \cite{hermes-pi0} the cross section for
$lp\rightarrow l'\pi^0 X$ was measured in dependence of the
azimuthal angle $\phi$, which is the angle between
lepton scattering plane and the plane defined by momentum ${\bf q}$
of virtual photon and momentum ${\bf P}_{\!h}$ of produced pion, see Fig. 2.

Denoting momentum of the target proton by $P$,
momentum of the incoming lepton by $l$ and
momentum of the outgoing lepton by $l'$,
the relevant kinematical variables --
center of mass energy square $s$,
four momentum transfer $q$,
invariant mass of the photon-proton system $W$,
$x$, $y$ and $z_h$ -- are defined as
\ba\label{notation-1}
        s   &:=& (P+l)^2 \;,\;\;\;
        q    :=  l-l'    \;,\;\;\;
        Q^2  :=  - q^2   \;,\;\;\;
        \nonumber\\
        W^2  &:=&(P+q)^2    = s(1-x) y + \Mn^2
        \nonumber\\
        x   &:=& \frac{Q^2}{2Pq}\;,\;\;\;
        y    :=  \frac{2Pq}{s}\;\;\;\mbox{and}\;\;\;
        z_h  :=  \frac{PP_h}{Pq\;} \;. \ea
In this notation the azimuthal asymmetry $A_{UL}^{\sin\phi}(x)$
measured by HERMES reads
\be\label{expl-1}
A_{UL}^{\sin\phi}(x) =
   \frac{\displaystyle\int\!\!\di y\,\di z_h\,\di\phi\,\sin\phi\left(
   \frac{1}{S^+}\,\frac{\di^4\sigma^+}{\di x\,\di y\,\di z_h \di\phi}-
   \frac{1}{S^-}\,\frac{\di^4\sigma^-}{\di x\,\di y\,\di z_h \di\phi}\right)}
        {\;\;\;\;\;\;\;\displaystyle
   \frac{1}{2}\int\!\!\di y\,\di z_h\di\phi\,\left(
   \frac{\di^4\sigma^+}{\di x\,\di y\,\di z_h\di\phi}+
   \frac{\di^4\sigma^-}{\di x\,\di y\,\di z_h\di\phi}\right)}\;\;.\ee
The subscript ``$U$'' reminds of the unpolarized beam, and
``$L$'' reminds of the longitudinally (with respect to the beam direction)
polarized proton target. $S^\pm$ denotes the proton spin,
where ``$^+$'' means polarization opposite to the beam direction.
When integrating over $y$ and $z_h$ one has to consider the experimental cuts
\be\label{exp-cuts}
    	W^2 > W_{\rm min}^2 = 4\,{\rm GeV}^2     , \;\;\;
    	Q^2 > Q^2_{\rm min}=1\,{\rm GeV}^2       , \;\;\;
    	0.2 <  y    < 0.85                       , \;\;\;
    	0.2 <  z_h  < 0.7                        \; . \ee

\paragraph{The azimuthal asymmetry.}
The cross sections entering the asymmetry $A_{UL}^{\sin\phi}$
Eq.(\ref{expl-1}) have been computed in Ref. \cite{Mulders:1996dh}
at tree-level up to order $1/Q$.
The denominator in Eq.(\ref{expl-1}) is the cross section for pion production
from scattering of unpolarized positrons on unpolarized target protons
\be\label{expl-den}
    \frac{1}{2}\left(
    \frac{\di^3\sigma^+}{\di x\,\di y\,\di\phi}+
    \frac{\di^3\sigma^-}{\di x\,\di y\,\di\phi}\right)=
    \frac{\di^3\sigma_{UU}}{\di x\,\di y\,\di\phi}\;\, .\ee
The numerator in Eq.(\ref{expl-1}) consists of two parts --
a longitudinal and a transverse part
with respect to the photon momentum ${\bf q}$
\be\label{expl-num}
     \frac{1}{S^+}\,\frac{\di^3\sigma^+}{\di x\,\di y\,\di\phi}
    -\frac{1}{S^-}\,\frac{\di^3\sigma^-}{\di x\,\di y\,\di\phi}
    =\frac{2}{S}\;\frac{\di^3\sigma_{UL}}{\di x\,\di y\,\di\phi}
    +\frac{2}{S}\;\frac{\di^3\sigma_{UT}}{\di x\,\di y\,\di\phi}\; . \ee
The cross sections are given by
\ba\label{expl-cross}
\frac{\di^4\sigma_{UU}}{\di x\,\di y\,\di z_h\di\phi}&=&
    \phantom{\sin\phi\,S_L\,}
    \frac{\alpha^2s}{Q^4}\;\biggl(1+(1-y)^2\biggr)
    \sum_{a}e_a^2\,xf_1^a(x)\, D_1^a(z_h) \nonumber\\
\frac{\di^4\sigma_{UL}}{\di x\, \di y\,\di z_h\di\phi}&=&
    \sin\phi\, S_L \frac{\alpha^2 s}{Q^4}\;\frac{\Mn}{Q}\;
    \frac{\,8(2-y)\sqrt{1-y\,}}{\la z_h\ra \sqrt{1+
    \la \Pnperp^2\ra/\la{\bf k}_{\!\perp}^2\ra}\;}
    \sum_a e_a^2\,x^3\!\!\int\limits_x^1\!\frac{\di\xi}{\xi^2}\,h_1^a(\xi)
    \,H_1^{\perp a}(z_h) \nonumber\\
\frac{\di^4\sigma_{UT}}{\di x\, \di y\,\di z_h\di\phi} &=&
    \sin\phi\,S_T
    \frac{\alpha^2 s}{Q^4}\;
    \frac{2(1-y)}{\la z_h\ra\sqrt{1+\la\Pnperp^2\ra/
    \la {\bf k}_{\!\perp}^2\ra}\;} \sum_a e_a^2 x\,h_1^a(x)\,
        H_1^{\perp a}(z_h) \;.\ea
In Eq.(\ref{expl-cross}) terms have been omitted which vanish after the
(weighted) integration over $\phi$, and pure twist-3 contributions have been
systematically neglected for reasons mentioned above
so that for $h_L$ entering $\sigma_{UL}$ the Wandzura-Wilczek type relation
$h_L(x)= 2x\int_x^1\di \xi ({h_1(\xi)}/{\xi^2})$ is hold
(see Ref. \cite{Mulders:1996dh} and Appendix).
A term proportional to $\widetilde{H}_1^\perp(z_h)$ is also neglected,
even though it contains a twist two contribution due to
$\widetilde{H}_1^\perp(z_h)=z_h({\di\;}/{\di z_h})H_1^\perp(z_h) + $
twist-3 \cite{Boglione:2000jk}. However the contribution of this term 
to $\sigma_{UL}$ is very small, see the Appendix.
$\la\Pnperp^2\ra$ and
$\la {\bf k}_{\!\perp}^2\ra=\la\Phperp^2\ra/\la z_h^2\ra$
are the mean square transverse momenta of quarks in the distribution and
fragmentation functions, respectively.
$S_L$ is the longitudinal, $S_T$ is the transverse component of
target spin $S$ with respect to the 3-momentum of the virtual photon
\be\label{A-spins-def}
    S_L = S\,\cos\theta_\gamma\simeq
          S\left(1-\frac{2\Mn^2x(1-y)}{s y}\right) \; ,\;\;\;
    S_T = S\,\sin\theta_\gamma\simeq
              S\;\sqrt{\displaystyle\frac{4\Mn^2x(1-y)}{s y}\;} \;,\ee
where $\theta_\gamma$ is the angle of virtual photon with respect to
incoming beam.

Assuming isospin symmetry and favoured fragmentation the following relations
hold
\ba\label{eq-favoured-frag}
D_1^{ u/\pi^+}\!  = D_1^{\bar d/\pi^+}\! =
D_1^{ d/\pi^-}\!  = D_1^{\bar u/\pi^-}\! &\gg&
D_1^{ d/\pi^+}\!  = D_1^{\bar u/\pi^+}\! =
D_1^{ u/\pi^-}\!  = D_1^{\bar d/\pi^-}\! \simeq 0 \nonumber\\
D_1^{ u/\pi^0}\!  = D_1^{\bar u/\pi^0}\! =
D_1^{ d/\pi^0}\!\;= D_1^{\bar d/\pi^0}\! \;&\;\mbox{and}\;&
D_1^{ u/\pi^+}\!  = D_1^{d/\pi^-}\! =   \frac{1}{2} D_1^{ u/\pi^0}
\eqdef D_1 \;, \ea
where the arguments $z_h$ are omitted.
The same relations hold for $H_1^\perp$.
Inserting Eq.(\ref{expl-cross}) and (\ref{eq-favoured-frag}) into
Eq.(\ref{expl-1}) for the azimuthal asymmetry $A_{UL}^{\sin\phi}$ yields
for the production of the pion
\ba\label{A-3}
    A_{UL}^{\sin\phi}(x,\pi)
&=&   \frac{1}{\la z_h\ra\sqrt{1+\la\Pnperp^2\ra/\la{\bf k}_{\!\perp}^2\ra}
      \;}\;\frac{\la H_1^\perp\ra}{\la D_1\ra}\nonumber\\
&\times&\Biggl(
      B_L(x)\;\frac{\sum_a^\pi e_a^2\,x^2 \int_x^1\!\di\xi\,h_1^a(\xi)/\xi^2}
      {\sum_{a'}^\pi e_{a'}^2\, f_1^{a'}(x)}
    + B_T(x)\;\frac{\sum_a^\pi e_a^2\, h_1^a(x)}
      {\sum_{a'}^\pi e_{a'}^2\,f_1^{a'}(x)\,}\Biggr) \, ,\ea
where $\sum_a^\pi$ means summation only over those flavours which contribute
to the favoured fragmentation into the specific pion
asymmetry, i.e. in the $\pi^0$ case e.g.
$$  \frac{\sum_a^{\pi^0}e_a^2\,h_1^a(x)}
         {\sum_{a'}^{\pi^0} e_{a'}^2\,f_1^{a'}(x)}
       = \frac{(4h_1^u+4h_1^{\bar u}+h_1^d+h_1^{\bar d})(x)}
              {(4f_1^u+4f_1^{\bar u}+f_1^d+f_1^{\bar d})(x)\;} \;.$$
The prefactors $B_L(x)$, $B_T(x)$ introduced in Eq.(\ref{A-3}) are given by
\be\label{A-prefactors-def}
    B_L(x) = \frac{\int\!\di y\,8(2-y)\,\sqrt{1-y}\;\cos\theta_\gamma\Mn/Q^5}
              {\int\!\di y\,(1+(1-y)^2)\,/\,Q^4} \;\;\;{\rm and}\;\;\;
    B_T(x) = \frac{\int\!\di y\,2(1-y)\,\sin\theta_\gamma/Q^4}
          {\int\!\di y\,(1+(1-y)^2)\,/Q^4}  \;\;. \ee

When integrating over $y\in[y_1(x),y_2(x)]$ one has to keep in mind that
$Q$, $\sin\theta_\gamma$ and $\cos\theta_\gamma$ are functions of
$x$ and $y$, according to Eq.(\ref{notation-1}) and Eq.(\ref{A-spins-def}).
The $x$-dependent integration range of variable $y$
is due to the experimental cuts Eq.(\ref{exp-cuts})
\be\label{exp-cuts-2}
    y_1(x) := \max\left(0.2,\;\frac{Q_{\rm min}^2}{sx},\;
              \frac{W_{\rm min}^2-\Mn^2}{s(1-x)}\right) \le y \le
    y_2(x) := 0.85\;. 
\ee
The implicit dependence of $h_1^a$, $f_1^a$, $H_1^{\perp a}$ and
$D_1^a$ on $y$ through $Q$ is neglected.
The distributions will be taken at the average value
$Q^2_{\rm av}=4\,{\rm GeV}^2$ close to the HERMES experiment.

\paragraph{Results.}
In the HERMES experiment $\la\Pnperp^2\ra\simeq
\la\Phperp^2\ra=\la z_h^2\ra\,\la{\bf k}_{\!\perp}^2\ra$ and
$\la z_h\ra=0.41$. Approximating $\la z_h^2\ra \simeq \la z_h\ra^2$ and using
the result Eq.(\ref{apower}), the overall prefactor in Eq.(\ref{A-3}) is
\be\label{A-B0-num-value}
    \frac{1}{\la z_h\ra\sqrt{1+\la\Pnperp^2\ra/\la{\bf k}_{\!\perp}^2\ra}
    \;}\;\frac{\la H_1^\perp\ra}{\la D_1\ra}
    = 0.12 \pm 0.04 \;.\ee
The error is due to the experimental error of the analyzing power
$\la H_1^\perp\ra/\la D_1\ra$ Eq.(\ref{apower}),
of which only the modulus is known.
Here we have chosen the positive sign, for which the analysis of azimuthal
asymmetries for $\pi^\pm$ gave evidence for \cite{Efremov:2000za}.
When using the DELPHI result Eq.(\ref{apower}) to explain the HERMES
experiment, we assume a  weak scale dependence of the analyzing power.
For $h_1^a(x)$ we take the results of the chiral quark-soliton model
\cite{h1-model} and for $f_1^a(x)$ the parametrization from Ref. \cite{GRV},
both LO-evolved to the average scale $Q_{\rm av}^2=4\,{\rm GeV}^2$.
  \begin{figure}[t!]
    \psfrag{x h(x)}{\boldmath\footnotesize\hspace{-0.5cm}
    ${\rm x\,h}_1^{\rm u}({\rm x}) \,\,{\rm vs.}\,\,
    {\rm x^3}\!\int_{\rm x}^1\!\di\xi{\rm h}_1^{\rm u}(\xi)/\xi^2$}
    \psfrag{t5}{\footnotesize\boldmath $A_{UL}^{\sin\phi}(x)$ for $\pi^0$}
    \psfrag{h1(x)}{\boldmath\footnotesize\hspace{-0.2cm}
    ${\rm x\, h}_1^{\rm u}({\rm x})$}
    \psfrag{hL(x)}{\boldmath\footnotesize\hspace{-0.5cm}
    ${\rm x^3}\!\int_{\rm x}^1\!\di\xi{\rm h}_1^{\rm u}(\xi)/\xi^2$}
  \begin{tabular}{cccccc}
    \hspace{-1cm} &
    \includegraphics[width=5.5cm,height=5.5cm]{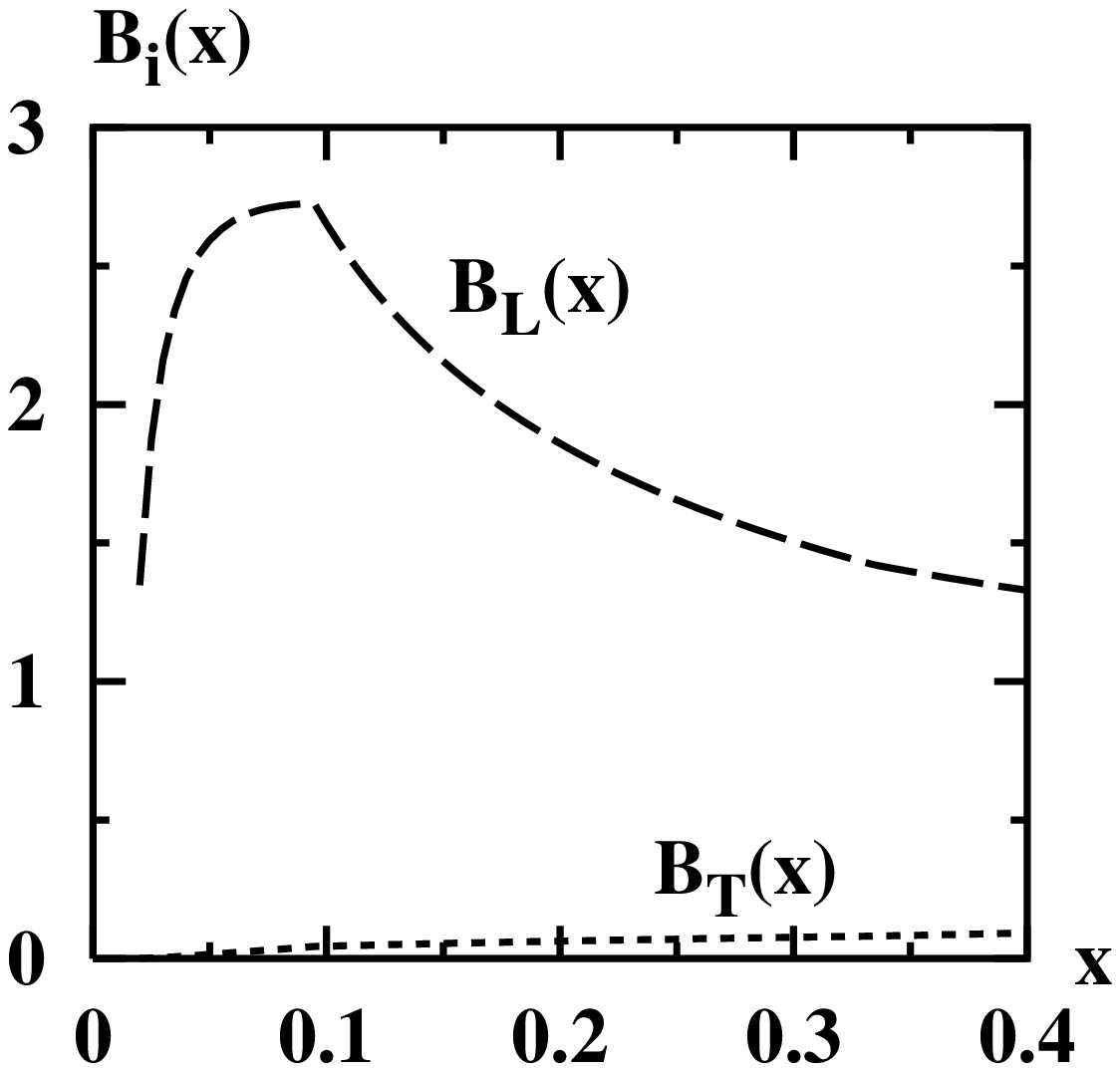} &
    \hspace{-1cm} &
    \includegraphics[width=5.5cm,height=5.5cm]{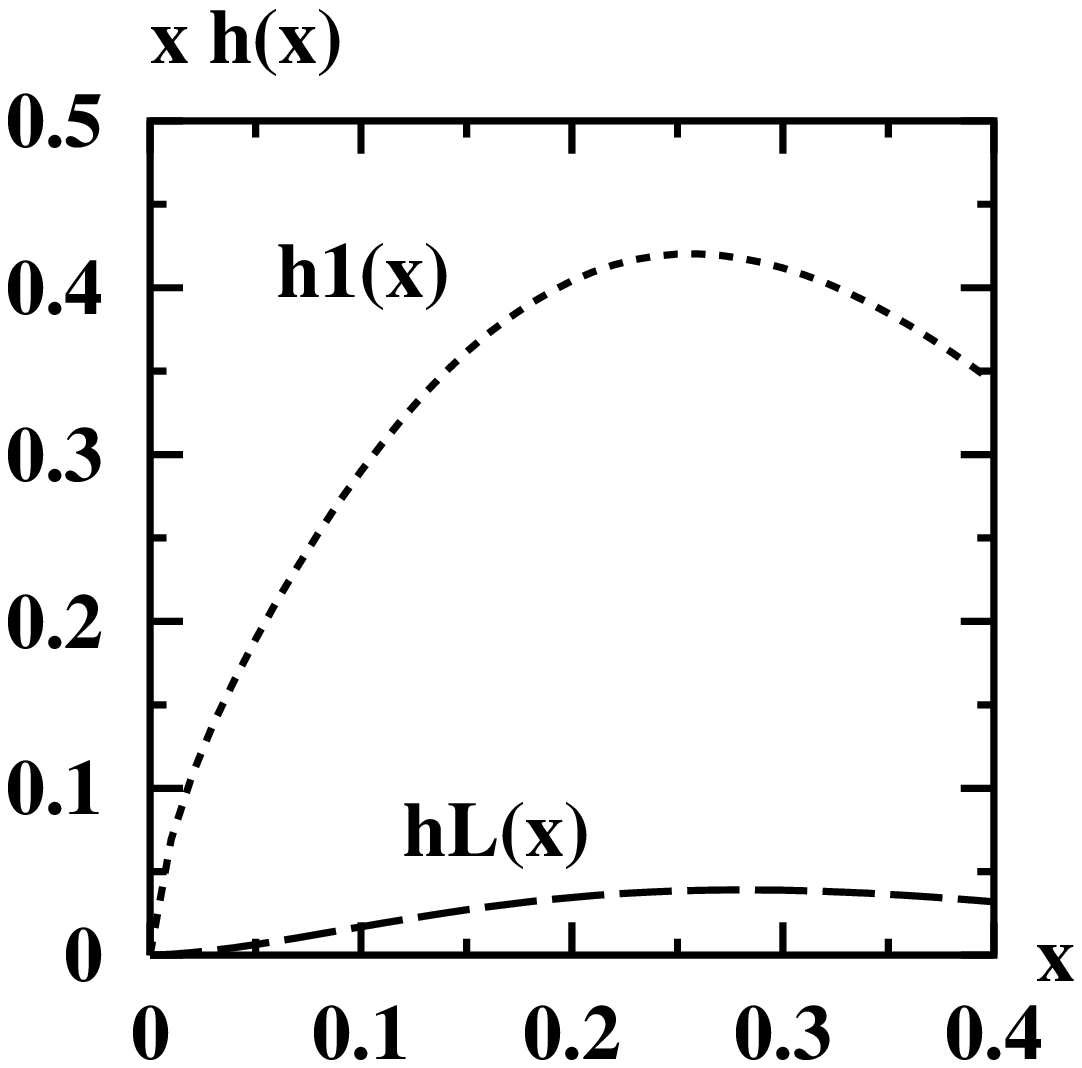} &
    \hspace{-1cm} &
    \includegraphics[width=5.5cm,height=5.5cm]{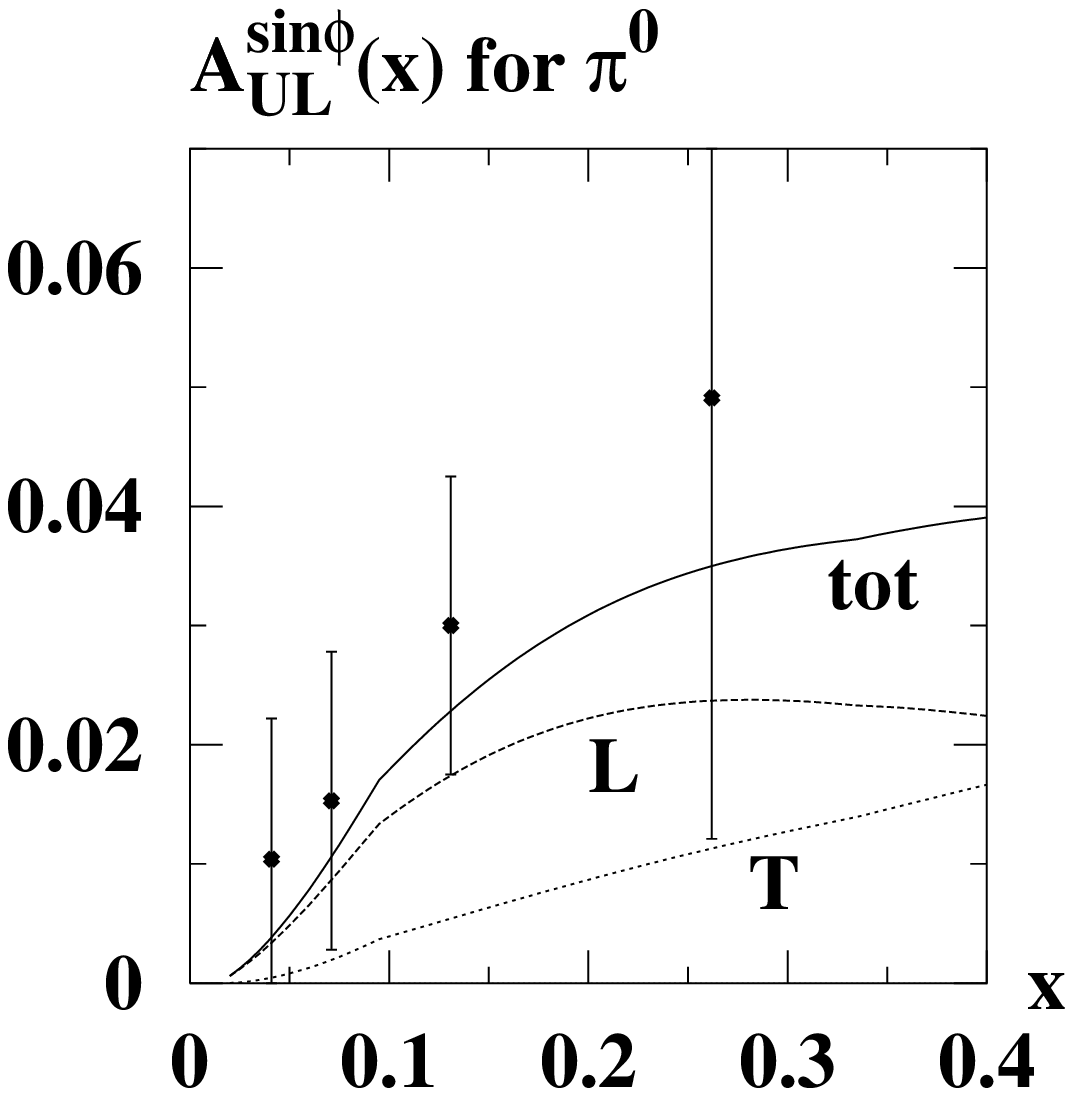}\cr
    \hspace{-1cm} & {\bf a} & \hspace{-1cm} & {\bf b} &
    \hspace{-1cm} & {\bf c}
  \end{tabular}
    \caption{{\bf a.} {\footnotesize\sl
    The prefactors $B_L(x)$ (dashed) and $B_T(x)$ (dotted line)
    -- as defined in Eq.(\ref{A-prefactors-def}) -- vs. $x$.
    Clearly $B_L(x)\gg B_T(x)$ for HERMES kinematics.} \newline
    Figure {\boldmath 3}: {\bf b.} {\footnotesize\sl
    $x^3\!\int_x^1\!\di\xi h_1^u(\xi)/\xi^2$ (dashed) and
    $xh_1^u(x)$ (dotted line) at $Q^2=4\,{\rm GeV}^2$ vs. $x$.
    One observes that
    $xh_1^u(x)\gg x^3\!\int_x^1\!\di\xi h_1^u(\xi)/\xi^2$.
    The situation is similar for other flavours.} \newline
    Figure {\boldmath 3}: {\bf c.} {\footnotesize\sl
    The contribution of longitudinal (L, dashed) and
    transverse (T, dotted) spin part to the total
    (tot, solid line) azimuthal $\pi^0$ asymmetry
    $A^{\sin\phi}_{UL}(x)$ and data from \cite{hermes-pi0} vs. $x$.}}
  \end{figure}

It is instructive to investigate how much the longitudinal spin (twist-3)
part and the transverse spin (twist-2) part contribute to the total azimuthal
asymmetry $A_{UL}^{\sin\phi}(x)$.
Comparing the $x$-dependent prefactors $B_L(x)$ and $B_T(x)$
Eq.(\ref{A-prefactors-def}), we note that $B_L(x) \gg B_T(x)$, see Fig. 3a.
This is due to the fact that $\cos\theta_\gamma\simeq 1$ appears in $B_L(x)$,
while in $B_T(x)$ we have $\sin\theta_\gamma ={\cal O}(\Mn/\sqrt{s})$
which is very small. However this effect is partially canceled by the fact
that $x^2\int_x^1\di y\,h_1^a(y)/y^2$, which contributes to the longitudinal
(twist-3) part, is much smaller than $h_1^a(x)$, which contributes to the
transverse (twist-2) part.
The results of the chiral quark-soliton model for $h_1^a(x)$ satisfy
$|x^2\int_x^1\di y\,h_1^a(y)/y^2|<0.1\,|h_1^a(x)|$ in the whole $x$ region.
In Fig. 3b this is demonstrated for the $u$ quark.
As a result the longitudinal and the transverse part give
-- with increasing $x$ -- comparably large contributions to the total
$A_{UL}^{\sin\phi}(x)$.
However, the longitudinal part gives always the
major contribution, see Fig. 3c.
The results shown in Fig. 3c correspond to the central value of the
numerical prefactor, Eq.(\ref{A-B0-num-value}).
For comparison data from Ref. \cite{hermes-pi0} are included
in Fig. 3c.

  \begin{figure}[t!]
  \begin{tabular}{cccccc}
    \hspace{-1cm} &
    \includegraphics[width=5.5cm,height=5.5cm]{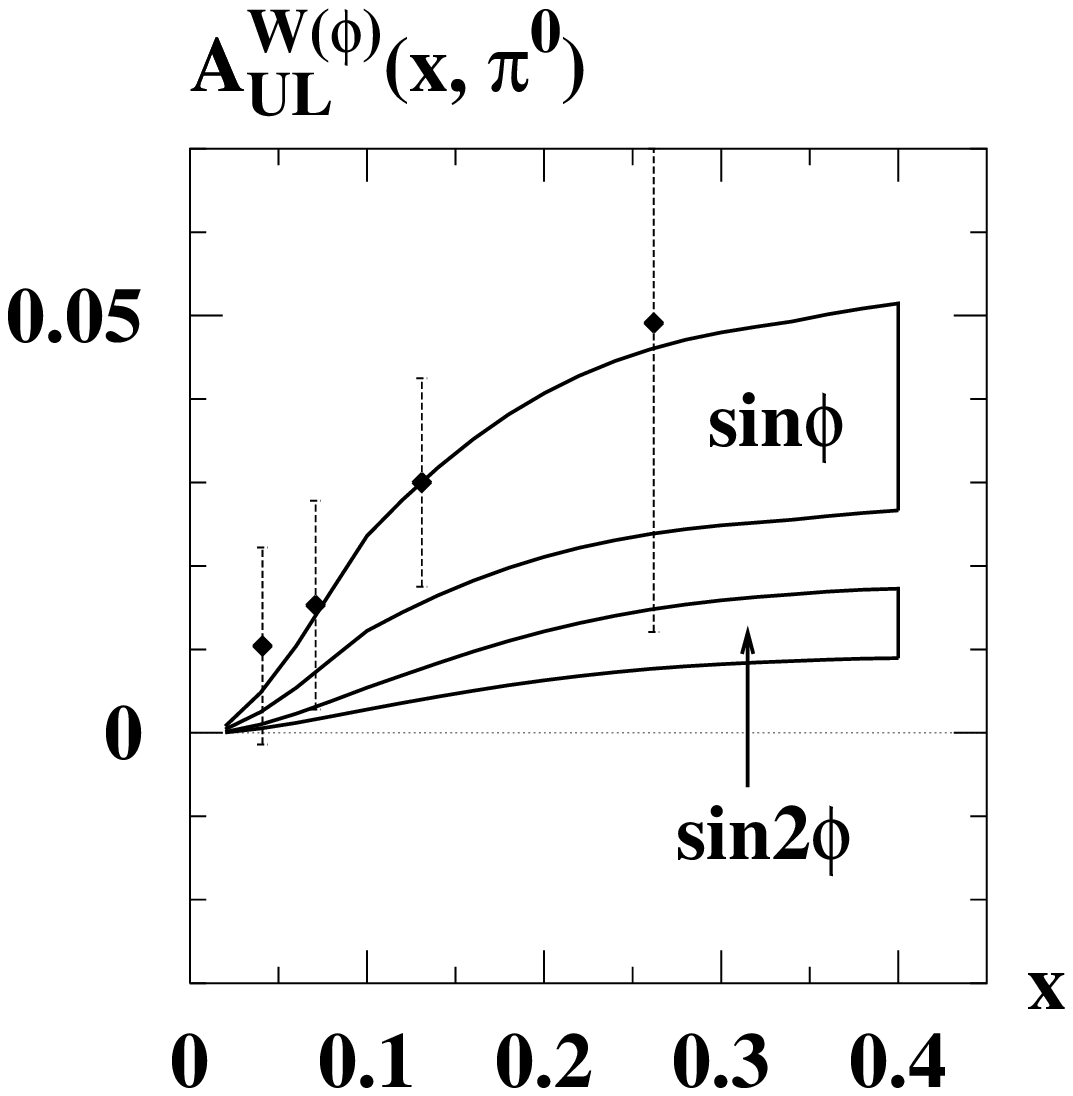} &
    \hspace{-1cm} &
    \includegraphics[width=5.5cm,height=5.5cm]{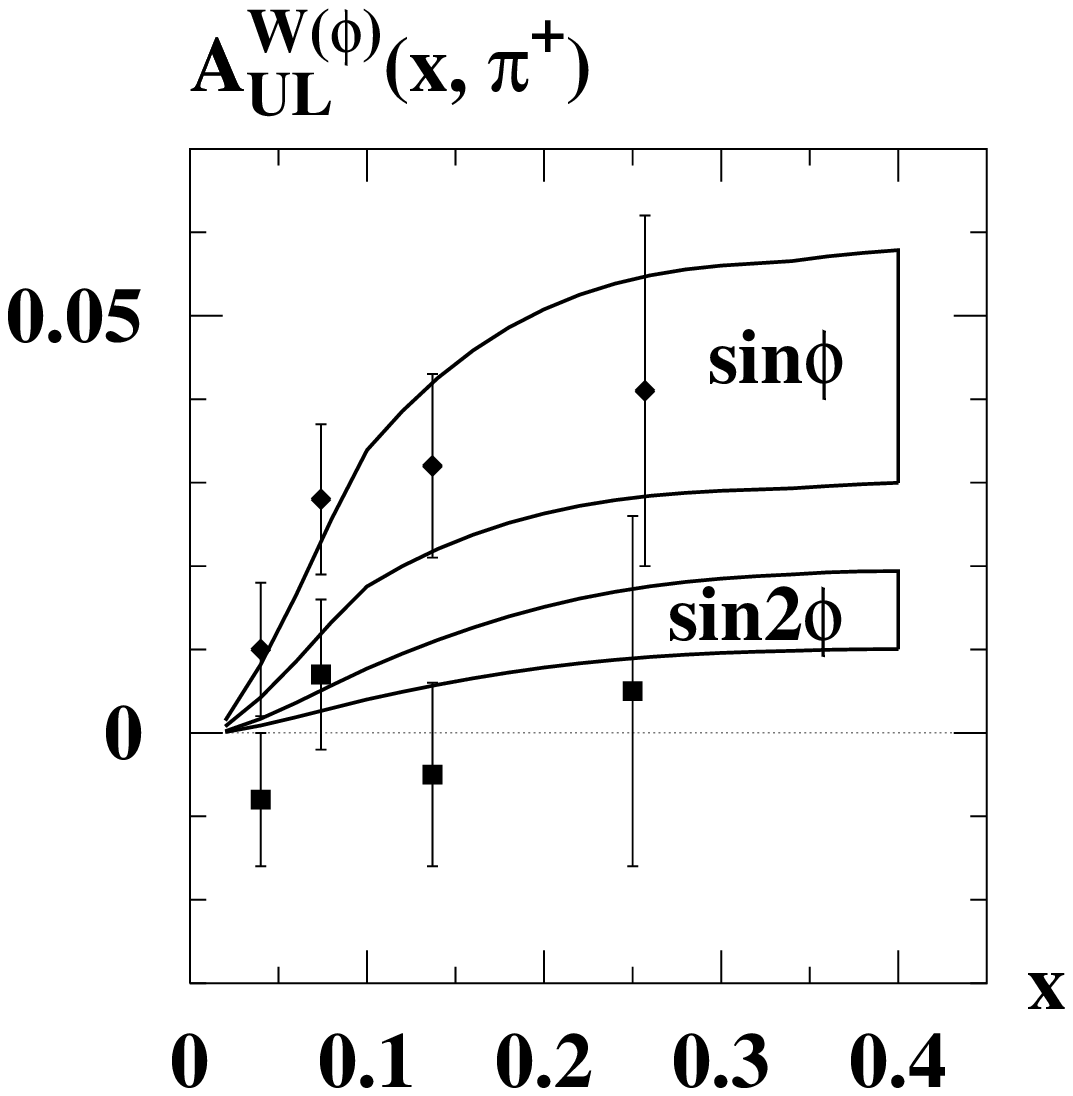} &
    \hspace{-1cm} &
    \includegraphics[width=5.5cm,height=5.5cm]{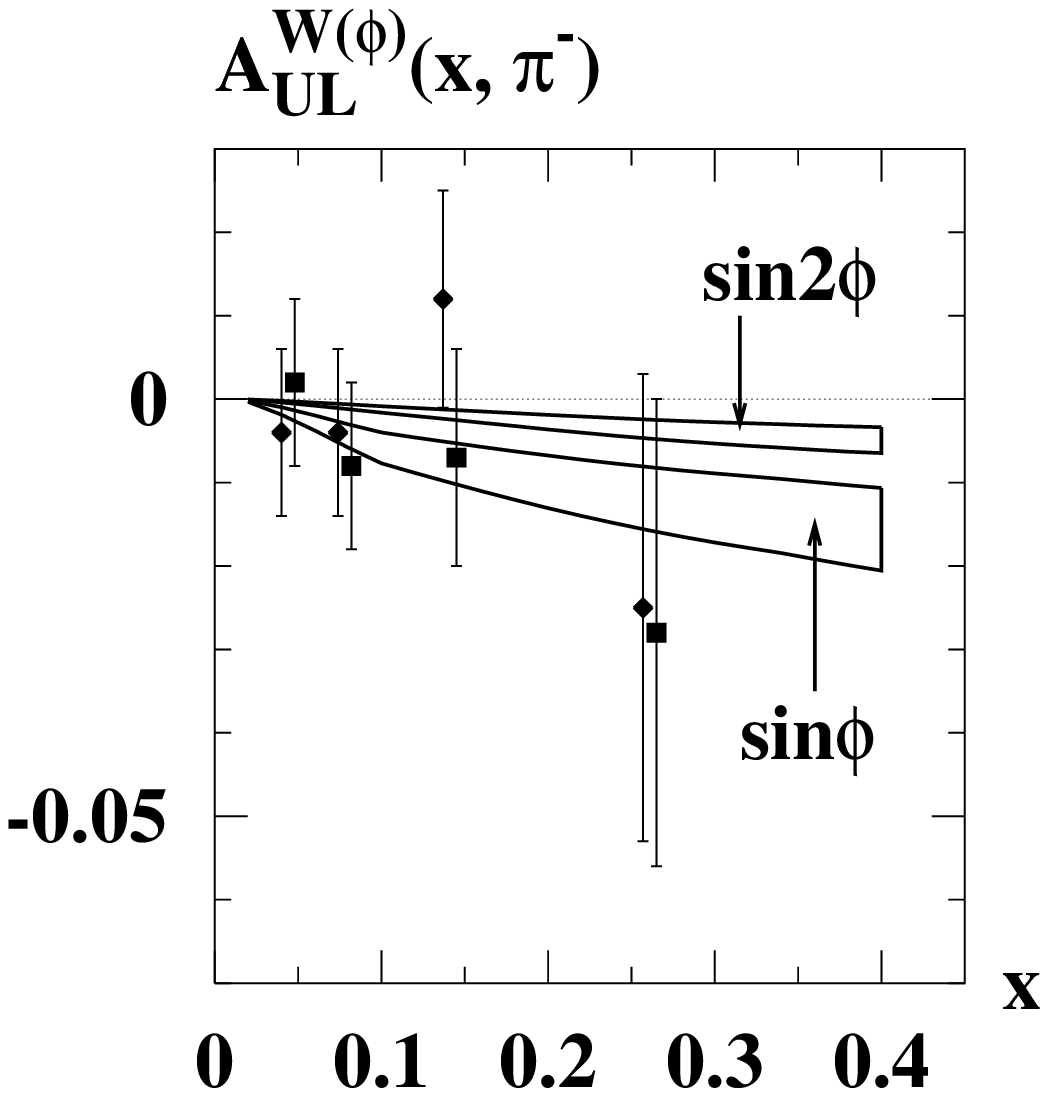}\cr
    \hspace{-1cm} & {\bf a} &
    \hspace{-1cm} & {\bf b} &
    \hspace{-1cm} & {\bf c}
  \end{tabular}
    \caption{\footnotesize\sl
    Azimuthal asymmetries $A_{UL}^{W(\phi)}(x,\pi)$
    weighted by $W(\phi)=\sin\phi$ and $\sin 2\phi$, respectively,
    for $\pi^0$ (a), $\pi^+$ (b) and $\pi^-$ (c) as function
    of $x$. The rhombuses denote data on $A_{UL}^{\sin\phi}(x,\pi)$,
    the squares data on $A_{UL}^{\sin 2\phi}(x,\pi)$ from
    Ref. \cite{hermes-pi0,hermes}. The enclosed areas correspond
    to the azimuthal asymmetries evaluated using the prediction of the
    chiral quark-soliton model for $h_1^a(x)$ and the DELPHI result for
    the analyzing power
    $\la H_1^\perp\ra/\la D_1\ra=(6.3\pm 2.0)\%$ \cite{todd},
    and take into account the statistical error of the analyzing power.}
  \end{figure}
Repeating the same steps for charged pions, we obtain the results shown
in Fig. 4.
In this figure we compare the HERMES data on $A_{UL}^{\sin\phi}(x)$
and $A_{UL}^{\sin 2\phi}(x)$ for $\pi^0$, $\pi^+$ and $\pi^-$
\cite{hermes-pi0,hermes} with the results which follow from our analysis.
The results shown here differ slightly from those obtained previously in
Ref. \cite{Efremov:2000za} since there the lower $y$-cut was taken to be
$y>0$, instead of $y>0.2$, see Eq.(\ref{exp-cuts}).

  \begin{table}[t!]
  \begin{center}
  \begin{tabular}{|l||c|c|}  \hline &&\\
  Asymmetry & $\chi$QSM \cite{h1-model} + DELPHI \cite{todd} &  HERMES exp.
  \cite{hermes-pi0,hermes} \\
  && \\ \cline{1-3}\hline\hline &&\\
  $A_{UL}^{\sin\phi}\;\;$
  \athree{\pi^0}{\pi^+}{\pi^-} &
  \athree{\phantom{-0}0.017 \pm 0.005}
         {\phantom{-0}0.024 \pm 0.008}
         {         - 0.0046 \pm 0.0015} &
  \athree{\phantom{-}0.019\pm0.007\pm0.003}
         {\phantom{-}0.022\pm0.005\pm0.003}
         {         - 0.002\pm0.006\pm0.004}\\
  && \\
  \cline{1-3}&&\\
  $A_{UL}^{\sin2\phi}\;\;$
  \athree{\pi^0}{\pi^+}{\pi^-} &
  \athree{\phantom{-}0.0044\pm0.0014}
         {\phantom{-}0.0063\pm0.0020}
         {         - 0.0011\pm0.0003} &
  \athree{\phantom{-}0.006\pm0.007\pm0.003}
         {         - 0.002\pm0.005\pm0.010}
         {         - 0.005\pm0.006\pm0.005}
  \\
  && \\ \hline
  \end{tabular}\end{center}\vspace{0.2cm}
      	\caption{\footnotesize\sl
      	The integrated azimuthal asymmetries $A_{UL}^{\sin\phi}$
    	and $A_{UL}^{\sin2\phi}$ for $\pi^+$, $\pi^0$ and $\pi^-$.
        $2^{\rm nd}$ column:
        Results obtained with the chiral quark-soliton model prediction
        for proton transversity distribution $h_1^a(x)$ \cite{h1-model}
    	and the DELPHI result for $H_1^\perp$ \cite{todd}.
    	The error is due to the statistical error of the DELPHI result,
        Eq.(\ref{apower}). \newline
        $3^{\rm rd}$ column: Experimental data from HERMES
        \cite{hermes-pi0,hermes}.}
  \end{table}
%
Finally, integrating the azimuthal asymmetries (numerator and denominator
separately) over the $x$-region covered by the HERMES experiment,
$0.023\le x\le 0.4$, we obtain the results for $A_{UL}^{\sin\phi}$ and
$A_{UL}^{\sin 2\phi}$ for $\pi^0$ and $\pi^\pm$ production which are
summarized in Table 1.
We conclude that the azimuthal asymmetries obtained with the chiral
quark-soliton model prediction for $h_1^a(x)$ \cite{h1-model}
combined with the DELPHI result for the analyzing power \cite{todd}
are consistent with experiment.

\section{Determining \boldmath$H_1^\perp(z_h)$}\label{sect-apower}

We used the DELPHI result for the analyzing power
$\la H_1^\perp\ra/\la D_1\ra$, Eq.(\ref{apower}), in order to explain the
HERMES experiment. When doing so we presumed that the analyzing power
varies  weakly with scale. This assumption can be questioned. Therefore
let us reverse the logic here, and use the HERMES results for the $\pi^0$ and
$\pi^+$ azimuthal asymmetries to estimate $H_1^\perp(z_h)/D_1(z_h)$.
For that we will use the chiral-quark soliton model prediction for
$h_1^a(x)$, and this will introduce a model dependence. However, since the
results of the model for known distribution functions agree
within (10 -- 20)\%
with parametrizations, we expect a similar ``accuracy'' for the
model prediction for $h_1(x)$.
With this in mind, the model dependence can be viewed as an additional
systematic error, which however is ``under control'' and of order
(10 - 20)\%.

From the HERMES data on $A_{UL}^{\sin\phi}(z_h)$ for $\pi^0$ and $\pi^+$
we obtain the results shown in Fig.~5.
The data can be described by the fit
\be\label{apower-fit}
    H_1^\perp(z_h) = a\,z_h\,D_1(z_h)\;\;\;\mbox{with}\;\;\;
        a = {\rm const} =  0.15 \pm 0.03
    \;\;. \ee
The error is the statistical error of the HERMES data.
One should keep in mind that there is also a systematical error of the
HERMES data (which varies with $z_h$), and a systematical error due to
the uncertainty of the theoretical calculation of $h_1^a(x)$.

  \begin{figure}[t!]
  \begin{center}
    \psfrag{H(z)/D(z)}{\hspace{-0.8cm}\footnotesize\bf\boldmath
    H$^\perp_1$(z$_{\rm h}$)/D$_1$(z$_{\rm h}$)}
    \psfrag{z}{\footnotesize\bf\boldmath z$_h$}
  \begin{tabular}{cccc}
    \hspace{-1cm} &
    \includegraphics[width=5.5cm,height=5.5cm]{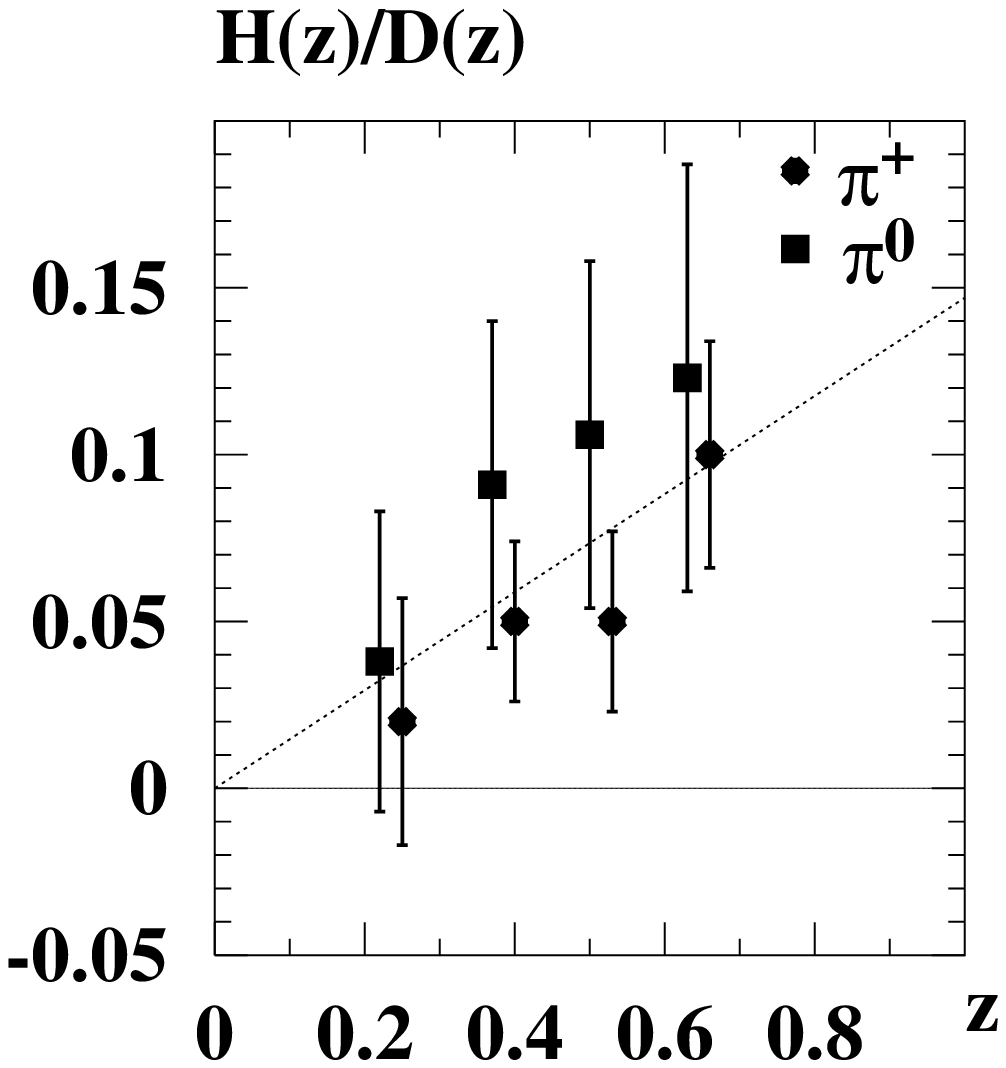} &
    \hspace{-1cm} &
    \includegraphics[width=5.5cm,height=5.5cm]{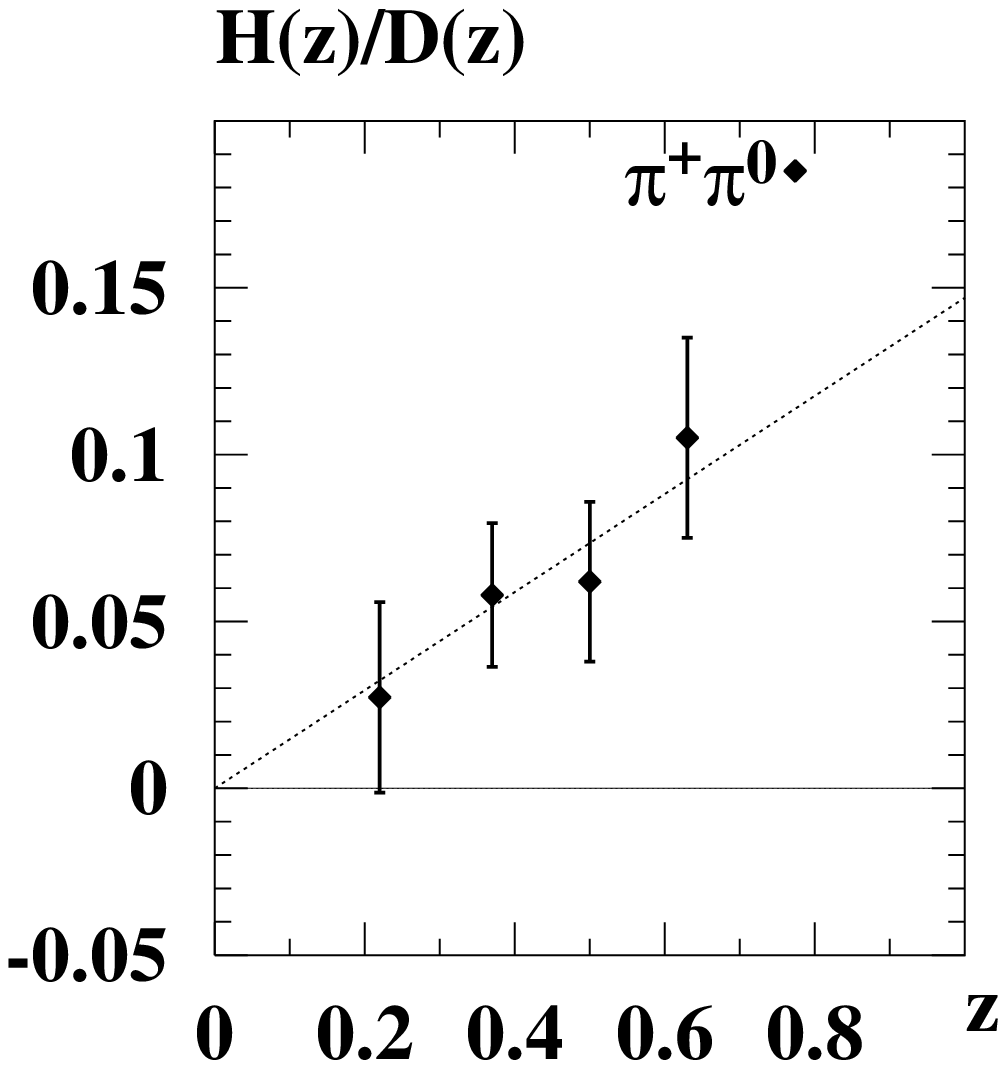} \\
    \hspace{-1cm} & {\bf a} & \hspace{-1cm} & {\bf b}
  \end{tabular}
  \end{center}
    \caption{{\bf a.} {\footnotesize\sl  $H_1^\perp(z_h)/D_1^\perp(z_h)$
    vs. $z_h$, as extracted from HERMES data
    \cite{hermes-pi0,hermes} on
    the azimuthal asymmetries $A_{UL}^{\sin\phi}(z_h)$ for $\pi^+$ and
    $\pi^0$ production using the prediction of the chiral quark-soliton
    model for $h_1^a(x)$ \cite{h1-model}. The error-bars are due to the
    statistical error of the data.}\newline
    Figure 5: {\bf b.} {\footnotesize\sl
        The same as Fig. 5a with data points from
    $\pi^+$ and $\pi^0$ combined.
    The line plotted in both figures is the best fit
    to the form $H_1^\perp(z_h)/D_1^\perp(z_h)= a\,z_h$
        with $a=0.15$.}}
  \end{figure}

Averaging over $z_h$ we obtain
\be\label{apower-HERMES+model}
    \frac{\la H_1^\perp\ra}{\la D_1\ra} = \cases{
    (5.8\pm 1.3\pm 0.8)\% & from HERMES $\pi^+$ data \cr
    (7.1\pm 2.6\pm 0.8)\% & from HERMES $\pi^0$ data \cr
    (6.1\pm 0.9\pm 0.8)\% & combined HERMES result.} \ee
Here the statistical and the systematical errors of the HERMES data are
considered. Again one should keep in mind an additional error of (10 - 20)\%
due to the uncertainty of theoretical prediction for $h_1^a(x)$.

Note that from the SMC data for the azimuthal asymmetry in the production of
charged hadrons in SIDIS off transversely polarized protons
\cite{bravardis99}, we obtain in this way the value
\be\label{apower-SMC+model}
    	\frac{\la H_1^\perp\ra}{\la D_1\ra} = (10 \pm 5)\%
	\;\;\;\;\mbox{from SMC data.}\ee
The results for the analyzing power from
HERMES Eq.(\ref{apower-HERMES+model}),
SMC    Eq.(\ref{apower-SMC+model}) and
DELPHI Eq.(\ref{apower}) are all consistent with each other.
This indicates that the scale dependence of $\la H_1^\perp\ra/\la D_1\ra$
might be indeed weak.

The observation that $H_1^\perp(z_h) \propto z_h\,D_1(z_h)$ -- if it will be
confirmed by future and more accurate data -- is physically very appealing.
The smaller the momentum fraction transferred from the parent parton to
the hadron, the less the produced hadron knows about the polarization of
the parton.
One should notice that such behaviour differs from those obtained in
simplest model calculation \cite{collins} and used by other
authors for explanation of HERMES asymmetries \cite{Oganessyan}.

\section*{Conclusions}

Using the result for $\la H_1^\perp\ra/\la D_1\ra$ from DELPHI \cite{todd}
and the chiral quark-soliton model prediction for the nucleon
transversity distribution $h_1^a(x)$ \cite{h1-model},
we obtain a good description of the azimuthal asymmetries
measured in the semi-inclusive $\pi^0$, $\pi^+$ and $\pi^-$ production
by HERMES \cite{hermes-pi0,hermes}.
The HERMES data suggest a strong flavour dependence of the transversity
distribution, a feature the chiral quark-soliton model results for
$h_1^a(x)$ successfully account for.
We stress that our description has no free adjustable parameters.

From the $z_h$-dependence of the HERMES data \cite{hermes-pi0,hermes} and
the chiral quark-soliton model prediction for $h_1^a(x)$ \cite{h1-model}
we extracted $H_1^\perp(z_h)/D_1(z_h)$ as function of $z_h$.
We find the Collins fragmentation function $H_1^\perp(z_h)$ proportional to
$z_h D_1(z_h)$ within (the large) error-bars in the $z_h$ region covered
in the experiment.

After averaging over $z_h$ we obtain a value for
$\la H_1^\perp\ra/\la D_1\ra$ very close to the DELPHI measurement.
This suggests that the scale dependence of the analyzing power
$\la H_1^\perp\ra/\la D_1\ra$ might be rather weak.
As an example of a similar behaviour of an asymmetry could serve the
ratio $A_1\propto G_1/F_2$ for which weak $Q^2$ dependence agrees with
the QCD evolution equations \cite{Kotikov:1997df}. Recently the evolution
equation for $H_1^\perp$ in the large $N_c$ limit was obtained
\cite{Henneman:2001ev}. This allows to make a similar investigation
for $ H_1^\perp/D_1$ which is under current study.

\medskip
{\small
We would like to thank M.~V.~Polyakov, K.~A.~Oganessyan
and O. Teryaev for fruitful discussions,
B.~Dressler for providing the evolution code, and D.~Hasch
from the HERMES collaboration for clarifying questions on experimental cuts.
A.E. is thankful to the Institute of Theoretical Physics II of
Ruhr University Bochum for warm hospitality.

\appendix
   \setcounter{equation}{0}
   \def\theequation{A.\arabic{equation}}
   \section{Azimuthal asymmetries}
   \label{Appendix}

\paragraph{Unpolarized cross section \boldmath$\sigma_{UU}$.}
The unpolarized differential cross section follows from Eq.(113) of
Ref. \cite{Mulders:1996dh}
\be\label{Muld-113}
\frac{\di^5\sigma_{UU}}{\di x\,\di y\,\di z_h\di^2{\bf P}_{\!\!\perp h}}
 = \frac{4\pi\alpha^2s}{Q^4}\sum_{a,\bar a}
e_a^2\Biggl\{\frac{1+(1-y)^2}{2}\,xf_1^a(x)D_1^a(z_h) + \dots\Biggr\}
\frac{{\cal G}(Q_\perp,R)}{z_h^2} \;. \ee
The dots denote terms which cancel out after the integration over $\phi$.
$Q_\perp=|{\bf q}_\perp|$ and
${\bf q}_\perp = -({\bf P}_{\!\!\perp h}/z_h)$.
The dependence of the distribution and fragmentation functions on transverse
quark momenta is assumed to be
\be\label{asym-2b+c}
    {\cal G}(Q_\perp,R) = \frac{R^2}{\pi}
    \exp\left(-Q_\perp^2R^2\,\right)\;\;\;{\rm with}\;\;\;
    \int\di^2{\bf P}_{\!\!\perp h}\; \frac{{\cal G}(Q_\perp,R)}{z_h^2}
    = 1 \; .\ee
After the integration over transverse momenta
$\di|{\bf P}_{\!\!\perp h}|\,|{\bf P}_{\!\!\perp h}|$, we obtain the spin
averaged cross section Eq.(\ref{expl-cross})
\ba\nonumber 
    \frac{\di^4\sigma_{UU}}{\di x\,\di y\,\di z_h\di\phi} =
    \frac{\alpha^2s}{Q^4}\;\left(1+(1-y)^2\right)\,
    \sum_{a,\bar a}e_a^2\,xf_1^a(x)\,D_1^a(z_h)\;.\ea

\paragraph{Longitudinal part \boldmath$\sigma_{UL}$.}
The part of $\sigma_{UL}$ which is proportional to $\sin\phi$ is given by
Eq.(115) of Ref. \cite{Mulders:1996dh}
\ba\label{asym-3b'}
    \frac{\di^5\sigma^{\sin \phi}_{UL}}
    {\di x\, \di y\,\di z_h\di^2 {\bf P}_{\!\!\perp h}\!\!\!}
&=& \sin\phi\, S_L\, \frac{4\pi\alpha^2 s}{Q^4}
    \,2(2-y)\sqrt{1-y\,}\,\frac{Q_\perp}{Q}\sum_a e_a^2\nonumber\\
&&  \times\;
      \Biggl\{
    \frac{R^6}{\Mn\la P_{\!\!\perp h}\ra R_{\Nu}^4R_h^4}
    \left[\frac{R_h^2-R_{\Nu}^2}{R^2}-Q^2_\perp R_h^2\right]
    xh_{1L}^{\perp a}(x) H_1^{\perp a}(z_h) \nonumber\\
&&  \;\;\;+\;
    \frac{\Mn\,R^2}{\la P_{\!\!\perp h}\ra R_h^2}\,x^2
    \widetilde{h}_L^a(x)H_1^{\perp a}(z_h) \nonumber\\
&&  \;\;\;-\;
    \frac{M_h^2 R^2}{\Mn\la P_{\!\!\perp h}\ra R_{\Nu}^2}
    \,xh_{1L}^{\perp a}(x) \frac{\widetilde{H}^a(z_h)}{z_h}
      \Biggr\}
    \frac{{\cal G}(Q_\perp;R)}{z_h^2} \; . \ea
Here $R^2=R_{\Nu}^2R_h^2/(R_{\Nu}^2+R_h^2)$.
Note the different normalization
\be\label{different-normalization}
    \frac{H_1^\perp}{M_h}
        \Biggr|_{\mbox{\footnotesize Ref. \cite{Mulders:1996dh}}}
    =\frac{H_1^\perp}{\la P_{\!\!\perp h}\ra}
        \Biggr|_{\mbox{\footnotesize here, Ref. \cite{Efremov:2000za}}}\;.\ee
Let us decompose $\sigma^{\sin \phi}_{UL}=\sigma^{\sin\phi}_{UL}[H_1^\perp]
+ \sigma^{\sin \phi}_{UL}[\widetilde{H}]$ and compute first the part
$\sigma^{\sin \phi}_{UL}[H_1^\perp]\propto H_1^\perp$ in Eq.(\ref{asym-3b'}).
Using the Wandzura--Wilczek type relation,
Eq.(C11) in Ref. \cite{Mulders:1996dh}
$$  h_L(x,{\bf P}_{\!\!\perp\Nu}^2) =
    - \frac{{\bf P}_{\!\!\perp\Nu}^2}{\Mn^2}
      \,\frac{h_{1L}^\perp(x,{\bf P}_{\!\!\perp\Nu}^2)}{x} +
      \widetilde{h}_L(x,{\bf P}_{\!\!\perp\Nu}^2) + {\cal O}(m_q/\Mn) \;,$$
and neglecting quark mass terms, we arrive at the relation
\be\label{asym-WWrel}
    {\bf P}_{\!\!\perp\Nu}^2\,h_{1L}^\perp(x,{\bf P}_{\!\!\perp\Nu}^2)
     = {\Mn^2}\,x\,\Biggl(\widetilde{h}_L(x,{\bf P}_{\!\!\perp\Nu}^2)
    - h_L(x,{\bf P}_{\!\!\perp\Nu}^2)\Biggr)\;.\ee
We have to reconsider the integration over the transverse quark momenta in
the target nucleon.
According to Eq.(D7) in Ref. \cite{Mulders:1996dh} the term containing
$h_{1L}^{\!\perp a}(x)$ in Eq.(\ref{asym-3b'}) arises from the convolution
\ba\label{asym-convolution}
    I\left[({\bf \hat h}{\bf P}_{\!\!\perp\Nu})\,({\bf P}_{\!\!\perp\Nu})^2
    \, h_{1L}^\perp H_1^\perp\right] = \frac{Q_T R^6}{R_{\Nu}^4R_h^4}
    \left[\frac{R_h^2-R_{\Nu}^2}{R^2}-Q^2_\perp R_h^2\right]
    I[h_{1L}^\perp H_1^\perp] && \nonumber\\
    {\rm where}\;\;\; I[h_{1L}^\perp H_1^\perp]\equiv h_{1L}^{\perp a}(x)
    H_1^{\perp a}(z_h)\frac{{\cal G}(Q_\perp;R)}{z_h^2} &&\!\!\!\!\! . \ea
If we insert the relation Eq.(\ref{asym-WWrel}) into the above convolution
Eq.(\ref{asym-convolution}) we obtain
\ba\label{asym-convol-2}
        I\left[({\bf \hat h}{\bf P}_{\!\!\perp\Nu})\,({\bf P}_{\!\!\perp\Nu})^2
    \, h_{1L}^\perp H_1^\perp\right]
 & = &  \frac{{\Mn^2}Q_T R^2}{R_h^2}\,x\left(h_L(x) - \widetilde{h}_L(x)\right)
    H_1^{\perp a}(z_h)\frac{{\cal G}(Q_\perp;R)}{z_h^2},
\ea
due to Eq.(D5) in Ref. \cite{Mulders:1996dh}. The result
Eq.(\ref{asym-convol-2}) we insert into the cross
section Eq.(\ref{asym-3b'})
and observe that the contribution of $\widetilde{h}_L^a(x)$ cancels out
exactly
\ba\label{asym-4g}
    \frac{\di^5\sigma^{\sin \phi}_{UL}[H_1^\perp]}
    {\di x\, \di y\,\di z_h\di^2 {\bf P}_{\!\!\perp h}\!\!\!}
& = &   \sin\phi\, \frac{4\pi\alpha^2 s S_L}{Q^4}
    \,2(2-y)\sqrt{1-y\,}
    \sum_a e_a^2\,x^2 h_L^a(x) H_1^{\perp a}(z_h) \nonumber\\
&&  \times\;
    \frac{Q_\perp}{Q}\,\frac{\Mn R^2}{\la P_{\!\!\perp h}\ra R_h^2}
        \frac{{\cal G}(Q_\perp;R)}{z_h^2} \; .\ea
In the next step we integrate Eq.(\ref{asym-4g}) over
$|{\bf P}_{\!\!\perp h}|\,\di|{\bf P}_{\!\!\perp h}|$. This yields
\be\label{asym-4h}
    \frac{\di^4\sigma^{\sin \phi}_{UL}[H_1^\perp]}
    {\di x\, \di y\,\di z_h\di\phi} =
       \sin\phi\, \frac{4\pi\alpha^2 s S_L\Mn}{Q^5}\,2(2-y)\sqrt{1-y\,}
           I_1 \sum_a e_a^2\,x^2 h_L^a(x) H_1^{\perp a}(z_h)\;,\ee
where
\ba
\label{integral-1}    I_1 &\equiv&
    \int\!\!\di|{\bf P}_{\!\!\perp h}|\;|{\bf P}_{\!\!\perp h}|\,
    \frac{Q_\perp R^2}{\la P_{\!\!\perp h}\ra R_h^2}
        \frac{{\cal G}(Q_\perp;R)}{z_h^2} 
    =   \frac{1}{2\pi\,\la z_h\ra}
    \;\frac{1}{\sqrt{1+\la {\bf P}_{\!\!\perp\Nu}^2\ra/
    \la {\bf P}_{\!h}^{\perp2}/z_h^2\ra}} \; .\ea
When performing the integral $I_1$ we made use of the definitions
\be\label{def-Qperp}
    \la Q_\perp\ra \equiv \int\!\!\di^2 {\bf Q}_{\perp}\;|{\bf Q}_{\perp}|
    \,{\cal G}(Q_\perp;R_h)  = \frac{\sqrt{\pi}}{2R_h}
    \; , \;\;\;
    \la {\bf Q}_{\perp}^2 \ra = \int\!\!\di^2 {\bf Q}_{\perp}\;
    {\bf Q}_{\perp}^2\,{\cal G}(Q_\perp;R_h) = \frac{1}{R_h^2} \; ,\ee
and analog definitions for $\la P_{\!N}^\perp\ra$ and
$\la {\bf P}_{\!N}^{\perp2}\ra$.
By means of Eq.(C19) in Ref. \cite{Mulders:1996dh}
-- where we neglect systematically current quark mass terms and the twist-3
contribution $\widetilde h_L$ --
\ba\label{asym-rel-a}
    h_L(x) &=& 2x\int\limits_x^1\di \xi \frac{h_1(\xi)}{\xi^2}
           +{\cal O}(\widetilde h_L)+{\cal O}(m_q/\Mn)\;,\ea
we finally arrive at the result quoted in Eq.(\ref{expl-cross})
\ba \nonumber 
    \frac{\di^4\sigma^{\sin \phi}_{UL}[H_1^\perp]}
    {\di x\, \di y\,\di z_h\di\phi}
    =
    \sin\phi\, S_L\, \frac{\alpha^2 s}{Q^4}\;\frac{\Mn}{Q}\;
    \frac{\,8(2-y)\sqrt{1-y\,}}{\la z_h\ra\sqrt{1+
    \la {\bf P}_{\!\!\perp\Nu}^2\ra/
    \la {\bf P}_{\!h}^{\perp2}/z_h^2\ra}}
    \sum_a e_a^2\,x^3 \int\limits_x^1\di \xi \frac{h_1(\xi)}{\xi^2}
    H_1^{\perp a}(z_h)\;.\ea

Next we turn to the contribution
$\sigma_{UL}[\widetilde{H}]\propto\widetilde{H}$, and show that we can
neglect it. Note that $\widetilde{H}$ is normalized analogously to
Eq.(\ref{different-normalization}).
After integration over $|{\bf P}_{\!\!\perp h}|\,\di|{\bf P}_{\!\!\perp h}|$
we obtain
\be\label{asym-sup1}
    \frac{\di^4\sigma^{\sin \phi}_{UL}[\widetilde{H}]}
    {\di x\, \di y\,\di z_h\di\phi\!\!\!}
    = -\sin\phi\, S_L\, \frac{4\pi\alpha^2 s}{Q^4}
    \,2(2-y)\sqrt{1-y\,}\,\frac{M_h^2 R_h^2}{Q\Mn R_\Nu^2}\,I_1\,
    \sum_a e_a^2 \,x h_{1L}^{\perp a}(x) \frac{\widetilde{H}^a(z_h)}{z_h}
     \; , \ee
with $I_1$ as defined in Eq.(\ref{integral-1}).
From Ref. \cite{Mulders:1996dh}, Eq.(C.15) and (C.19), we obtain the relation
\be
    h_{1L}^{\perp a}(x) = -\,\frac{\Mn^2}{\la{\bf P}_{\!\perp N}^2\ra}\,x
    \left(2x\int\limits_x^1\!\di\xi\,\frac{h_1^a(\xi)}{\xi^2}+\dots\right)
    \;,\ee
where the dots denote twist-3 terms and contributions proportional
to current quark masses, which we neglect. We also use the relation
\cite{Boglione:2000jk}
\be
    \frac{\widetilde{H}^a(z_h)}{z_h} = \frac{\di\;}{\di z_h}\Biggl(
    z_h H_1^{\perp a}(z_h)\Biggr)+\dots \;,\ee
where we neglect consistently a twist-3 contribution,
and obtain
\ba
    \frac{\di^4\sigma^{\sin \phi}_{UL}[\widetilde{H}]}
    {\di x\, \di y\,\di z_h\di\phi\!\!\!}
    &=& \sin\phi\, S_L\, \frac{4\pi\alpha^2 s}{Q^4}
    \,2(2-y)\sqrt{1-y\,}\,
    \frac{\Mn}{Q} \left(M_h^2 R_h^2 \right)\;\nonumber\\
    &\times&
    \frac{1}{\sqrt{1+\la{\bf P}_{\!\!\perp N}^2\ra/
    \la {\bf P}_{\!\!\perp h}^2/z_h^2\ra}}
    \sum_a e_a^2 \,x^3 \int\limits_x^1\!\di\xi\,\frac{h_1^a(\xi)}{\xi^2}
    \;\frac{\di\;}{\di z_h}\Biggl(z_h H_1^{\perp a}(z_h)\Biggr)\;. \ea
Thus, using the relations Eq.(\ref{eq-favoured-frag}) and the
definition Eq.(\ref{def-Qperp}), we see that
\be
    \frac{\di^4\sigma^{\sin \phi}_{UL}[\widetilde{H}]}
    {\di x\, \di y\,\di z_h\di\phi\!\!\!}
    =
    \left(\frac{\pi\la z_h\ra^2M_h^2}{4\la P_{\!\!\perp h}\ra^2}\right)\;
    \frac{\frac{\di\;}{\di z_h}(z_h H_1^{\perp}(z_h))}{H_1^{\perp}(z_h)}
    \;\cdot\;\frac{\di^4\sigma^{\sin \phi}_{UL}[H_1^\perp]}
    {\di x\, \di y\,\di z_h\di\phi\!\!\!}
    \ll
    \frac{\di^4\sigma^{\sin \phi}_{UL}[H_1^\perp]}
    {\di x\, \di y\,\di z_h\di\phi\!\!\!}\;,\ee
not only due to
$\pi\la z_h\ra^2M_h^2/4\la P_{\perp h}\ra^2\sim0.1$ in the HERMES experiment.
Also $\frac{d}{d z_h}(z_h H_1^\perp(z_h))/H_1^\perp(z_h)
= z_h \frac{d}{d z_h}(\ln z_h H_1^\perp(z_h))$ is small
provided $H_1^\perp(z_h)$ is a smooth function for $0.2\le z_h\le 0.7$.

\paragraph{Transverse part \boldmath$\sigma_{UT}$.}
According to Eq.(116) in Ref. \cite{Mulders:1996dh} the only term which is
non-zero after the ($\sin\phi$-weighted) integration over $\phi$ reads
\be\label{app-UT-1}
    \frac{\di^5\sigma^{\sin \phi}_{UT}}
    {\di x\,\di y\,\di z_h\di^2{\bf P}_{\!\!\perp h}\!\!\!} =
    - \sin(\phi+\phi_s)S_T\,\frac{4\pi\alpha^2 s}{Q^4}\,(1-y)\,
    \frac{Q_\perp R^2}{\la P_{\!\perp h}\ra R_h^2}
    \sum_a e_a^2 xh_1^a(x) H_1^{\perp a}(z_h)\;
    \frac{{\cal G}(Q_\perp; R)}{z_h^2} \;,
\ee
with $\phi_s=-\pi$ for the longitudinally polarized target in the HERMES
experiment. After the integration over transverse momenta
we obtain the result quoted in Eq.(\ref{expl-cross})
\ba \nonumber 
    \frac{\di^4\sigma^{\sin \phi}_{UT}}
    {\di x\,\di y\,\di z_h\di\phi} =
    \sin\phi\, S_T\,\frac{\alpha^2 s}{Q^4}\,\,
    \frac{2(1-y)}{\la z_h\ra
    \sqrt{1+\la z_h^2\ra \la {\bf P}_{\!\!\perp\Nu}^2\ra/
    \la {\bf P}_{\!\!\perp h}^2\ra}}
    \sum_a e_a^2 xh_1^a(x) H_1^{\perp a}(z_h) \;.
\ea


\end{document}